\documentclass[12pt,twoside]{article}
\usepackage{amsmath}
\newcommand{\rd}[1]{\mathop{\mathrm{d}#1}}
\newcommand{\tr}{\mathop{\mathrm{tr}}}

\newcommand{\fract}[2]{{\textstyle\frac{#1}{#2}}}
\newcommand{\grad}{\vec\nabla}

\newcommand{\nA}{non-Abelian}
\newcommand{\CS}{Chern-Simons}
\newcommand{\CSt}{Chern-Simons term}
\newcommand{\Cpr}{Clebsch pa\-ra\-me\-ter\-iza\-tion}

\newcommand{\pr}{para\-me\-ter\-iza\-tion}
\newcommand{\prd}{para\-me\-ter\-ized}

\newcommand{\mn}{{\mu\nu}}
\newcommand{\pp}[1]{\partial_{#1}}

\newcommand{\gdg}{g^{-1} \rd g}
\newcommand{\A}{\mathcal A}

\newcommand{\numeq}[2]{\begin{equation}
#2
\label{#1}
\end{equation}}
\newcommand{\refeq}[1]{(\ref{#1})}

\let\vec\boldsymbol
\let\eps\varepsilon
\let\epsilon\varepsilon
\let\phi\varphi
\let\hat\widehat

\begin{document}
 
\title{Descendants of the Chiral Anomaly}
\author{R. Jackiw\\
\small\it Center for Theoretical Physics\\ 
\small\it Massachusetts Institute of Technology\\ 
\small\it Cambridge, MA 02139-4307}
\date{\small Dirac Medalist Meeting, Trieste, Italy, November 2000\\
MIT-CTP\#3051}
\maketitle

\abstract{\noindent
\CSt s are well-known descendants of chiral anomalies, when the latter are presented 
as total derivatives. Here I explain that also \CSt s, when defined on a 3-manifold,
may be expressed as total derivatives.}
\bigskip

\pagestyle{myheadings}
\markboth{\small {\it R. Jackiw}}{\small  Descendants of the Chiral Anomaly}
\thispagestyle{empty}

The axial anomaly, that is, the departure from transversality of the 
correlation function for fermion vector, vector, and axial vector currents, involves
${}^*\!FF$, an expression constructed from the gauge fields to which the fermions
couple. Specifically, in the Abelian case one encounters
\numeq{eq1}{
{}^*\!F^\mn  F_\mn = \fract12 \eps^{\mn\alpha\beta} F_\mn F_{\alpha\beta} =
-4
\vec E\cdot \vec B
 }
where $F_\mn$ is the covariant electromagnetic tensor 
\begin{subequations}
\numeq{eq2a}{
F_\mn = \pp\mu A_\nu - \pp \mu A_\nu
 }
while $\vec E$ and $\vec B$ are the electric and magnetic fields
\numeq{2a}{
 E^i = F^{io}\ ,\quad
 B^i = -\fract12 \eps^{ijk} F_{jk}\ .
}
\end{subequations}
The non-Abelian generalization reads
\numeq{eq3}{
{}^*\! F^{\mn a} F_\mn^a   = \fract12 \eps^{\mn\alpha\beta} F_\mn^a
F_{\alpha\beta}^a 
}
where $F_\mn^a$ is the Yang-Mills gauge field strength 
\numeq{eq4}{
F_\mn^a = \pp\mu A_\nu^a - \pp \nu A_\mu^a + f^{abc} A_\mu^b A_\nu^c
}
and $a$ labels the components of the gauge group, whose structure constants
are~$f^{abc}$. 

The quantity $ {}^*\!FF$ is topologically interesting. Its integral over 4-space is
quantized, and measures the topological class (labeled by integers) to which the
vector potential~$A$ belongs. Consequently, the  integral of $ {}^*\!FF$ is a
topological invariant and we expect that, as befits a topological invariant, it should be
possible to present
$ {}^*\!FF$ as a total derivative, so that its 4-volume integral becomes converted by
Gauss' law into a surface integral, sensitive only to long distance, global properties of
the gauge fields. That a total derivative form for
${}^*\!FF$ indeed holds is seen when $F_\mn$ is expressed in terms of potentials. In
the Abelian case, we use \refeq{eq2a} and find immediately  
\numeq{eq5}{
\fract12  {}^*\!F^\mn F_\mn  = \pp \mu \bigl(\eps^{\mu\alpha\beta\gamma}
A_\alpha \pp\beta A_\gamma \bigr)\ .
 }
For non-Abelian fields, \refeq{eq4} establishes the result we desire:
\numeq{eq6}{
\fract12 {}^*\!F^{\mn a} F_\mn^a   = \pp\mu \eps^{\mu\alpha\beta\gamma}
\bigl( A_\alpha^a\pp\beta A_\gamma^a + \fract13 f^{abc} A_\alpha^a A_\beta^b
A_\gamma^c
\bigr)\ .
}
The quantities whose divergence gives ${}^*\!FF$ are called \CS\ terms. By
suppressing one dimension they become naturally defined on a 3-dimensional
manifold (they are 3-forms), and we are thus led to consider the \CS\ terms in their
own right~\cite{ref7}:
\begin{align}
\mathrm{CS}(A) &= \eps^{ijk} A_i \pp j A_k & &\text{(Abelian)}\label{eq7}\\
\mathrm{CS}(A) &= \eps^{ijk}\bigl( A_i^a \pp j A_k^a +\fract13 f^{abc} A_i^a A_j^b
A_k^c \bigr)
 & &\text{(non-Abelian).}\label{eq8}
\end{align}

The 3-dimensional integral of these quantities is again topologically interesting. When
the
\nA\
\CSt\ is evaluated on a pure gauge, \nA\ vector potential
\numeq{eq9}{
A_i = g^{-1} \pp i g
} 
the 3-dimensional volume integral of $\mathrm{CS}(g^{-1} \partial  g)$ measures the
topological class (labeled by integers) to which the group element $g$ belongs. The
integral in the Abelian case -- the case of electrodynamics -- is called the magnetic
helicity
$\int
\rd{^3 r} \vec A\cdot
\vec B$,
$\vec B = \grad
\times \vec A$, and measures the linkage of magnetic flux lines. An analogous
quantity arises in fluid mechanics, with the local fluid velocity $\vec v$ replacing
$\vec A$, and the vorticity $\vec\omega = \grad\times\vec v$ replacing~$\vec B$. Then
the integral $\int \rd{^3 r} \vec v\cdot \vec\omega$ is called kinetic
helicity~\cite{ref8}.

I shall not review here the many uses to which the \CSt s, Abelian and \nA,
introduced in~\cite{ref7}, have been put. The applications range from the
mathematical characterizations of knots to the physical descriptions of electrons in
the quantum Hall effect~\cite{ref9}, vivid evidence for the deep significance of the
\CS\ structure and of its antecedent, the chiral anomaly. 

Instead, I pose the following question: Can one write the \CSt\ as a total derivative,
so that (as befits a topological quantity) the spatial volume integral becomes a surface
integral? 
 An argument that this should be possible is the
following: The
\CSt\ is a 3-form on 3-space, hence it is maximal and its exterior derivative vanishes
because there are no 4-forms on 3-space. This establishes that on 3-space  the \CSt\
is closed, so one can expect that it is also exact, at least locally, that is, it can be
written as a total derivative. 
Of course such a representation for the \CSt\ requires expressing the potentials  in
terms of ``prepotentials'', since the formulas \refeq{eq7}, \refeq{eq8} in terms of
potentials show no evidence of derivative structure. 
[Recall that the total derivative formulas \refeq{eq5}, \refeq{eq6} for the axial
anomaly also require using potentials to express~$F$.]

There is a physical, practical reason for wanting the Abelian \CSt\  to be a total
derivative. It is known in fluid mechanics that there exists an obstruction to
constructing a Lagrangian for Euler's fluid equations, and this obstruction is just the
kinetic helicity  \hbox{$\int \rd{^3 r} \vec v\cdot \vec\omega$}, that is, the volume
integral of the Abelian \CSt, constructed from the velocity 3-vector~$\vec v$. This
obstruction is removed when the integrand is a total derivative, because then the
kinetic helicity volume integral is converted to a surface integral by Gauss' theorem.
When the integral obtains contributions only from a surface,  the obstruction
disappears from the 3-volume, where the fluid equation acts~\cite{ref10}. 

It is easy to show that the  Abelian \CSt\ can be presented as a total derivative. We
use the \Cpr\ for  a 3-vector~\cite{ref11}:
\numeq{eq10}{
\vec A = \grad\theta + \alpha\grad\beta\ .
}
This nineteenth century \pr\ of a 3-vector $\vec A$ in terms of the prepotentials
($\theta$, $\alpha$, $\beta$) is an alternative to the usual transverse/longitudinal
\pr. In modern language it is a statement of Darboux's
theorem that the 1-form $A_i \rd{r^i}$ can be written as $\rd \theta + \alpha
\rd\beta$~\cite{ref12}. With this
\pr\ for $\vec A$, one sees that the Abelian \CSt\ indeed is  a total derivative:
\begin{align}
\mathrm{CS}(A) &= \eps^{ijk} A_i \pp j A_k\label{eq11}\\
&=  \eps^{ijk} \pp i \theta\pp j \alpha \pp k \beta\nonumber\\
&= \pp i \bigl( \eps^{ijk} \theta\pp j \alpha \pp k \beta\bigr)\ .\nonumber
\end{align}

When the \Cpr\ is employed for $\vec v$ in the fluid dynamical context, the
situation is analogous to the force law in electrodynamics. While the Lorentz equation
is written in terms of field strengths, a Lagrangian formulation needs potentials from
which the field strengths are reconstructed. Similarly, Euler's equation involves the
velocity vector~$\vec v$, but in a Lagrangian for this equation the velocity must be
parameterized in terms of the  prepotentials $\theta$, $\alpha$, and~$\beta$. 

In a natural generalization of the above, we ask whether a \nA\ vector potential
can also be \prd\ in such a way that the \nA\ \CSt~\eqref{eq8} becomes a total
derivative. We have answered this question affirmatively and we have found
appropriate prepotentials that do the job~\cite{ref10,ref13,ref14}.

In order to describe our \nA\ construction, we first revisit the Abelian problem. As
we have stated, the solution in the Abelian case is immediately provided by the
\Cpr~\refeq{eq10}. However, finding the \nA\ generalization requires an indirect
construction, which we first present for the Abelian case.

Although in the Abelian case we are concerned with U(1) potentials, we begin by
considering a bigger group SU(2), which contains our group of interest~U(1).  Let~$g$
be a group element of SU(2) and construct a pure-gauge SU(2) gauge potential
\numeq{eq12}{
{\A} = g^{-1} \rd g\ . 
}
We know that $\tr(\gdg)^3$ is a total derivative~\cite{ref7}; indeed, its 3-volume
integral  measures the topological winding number of~$g$ and therefore can be
expressed as a surface integral, as befits a topological quantity. The separate SU(2)
component potentials $\A^a$ can be projected from~\refeq{eq12} as
\numeq{eq13}{
\mathcal A^a= i\tr \sigma^a \gdg\ , \quad
\mathcal A = \mathcal A^a \sigma^a/2i
}
and
\begin{align}
-\fract23 \tr(\gdg)^3 &= \frac1{3!} \eps^{abc} \mathcal A^a \mathcal A^b \mathcal
A^c \nonumber\\
&= \A^1 \A^2 \A^3\ .
\label{eq14} 
\end{align}
Moreover, since $\A^a$ is a pure gauge, it satisfies
\numeq{eq15}{
\rd{\A^a} = -\fract12 \eps^{abc} \A^b \A^c \ .
}
Next define an Abelian vector potential~$A$ by projecting one component of $\gdg$
\numeq{eq16}{
A = i \tr \sigma^3 \gdg = \A^3\ .
}
Note that $A$ is \emph{not} an Abelian pure gauge $\grad\times\vec A = \vec B
\neq0$. It now follows from~\refeq{eq15} that 
\begin{align}
\vec A\cdot\vec B \rd{^3 r} &= A\rd A = \A^3\rd{\A^3} =
-\A^1\A^2\A^3\nonumber\\
&= \fract23 \tr(\gdg)^3\ .
\label{eq17}
\end{align}
The last equality is a consequence of~\refeq{eq14} and shows that the Abelian \CSt\
is proportional to the winding number density of the \nA\ group element, and
therefore is  a total derivative. Note that the projected formula~\refeq{eq16} involves
three arbitrary functions -- the three parameter functions of the SU(2) group --
which is the correct number needed to represent an Abelian vector potential in
3-space.

It is instructive to see how this works explicitly. The most general SU(2) group
element reads $\exp (\sigma^a\omega^a / 2i)$. The three functions $\omega^a$ are
presented as $\hat\omega^a \omega$, where  $\hat\omega^a$ is a unit SU(2)
3-vector and $\omega$ is the magnitude of  $\omega^a$. The unit vector may be
\prd\ as 
\begin{subequations}
\numeq{eq17a}{
\hat\omega^a = (\sin\Theta\cos\Phi, \sin\Theta\sin\Phi,\cos\Theta)
}
where $\Theta$ and $\Phi$ are functions on 3-space, as is~$\omega$. A simple
calculation shows that 
\numeq{eq17b}{
\gdg = \frac{\sigma^a}{2i} \bigl(
\hat\omega^a\rd\omega + \sin\omega\rd{\hat\omega^a} - 
(1-\cos\omega)\eps^{abc}\, \hat\omega^b  \rd{\hat\omega^c} 
\bigr)
}
\end{subequations}
\begin{align}
A &= \tr i\sigma^3\gdg\nonumber\\
  &= \cos\Theta\rd\omega -\sin\omega\sin\Theta\rd\Theta - 
(1-\cos\omega) \sin^2\Theta\rd\Phi\label{eq18}\\
A\rd A&= -2\rd{(\omega-\sin\omega)}\rd{(\cos\Theta)}\rd\Phi = \rd\Omega
\label{eq19}\\
\Omega &= -2\Phi\rd{(\omega-\sin\omega)}\rd{(\cos\Theta)}\ .\label{eq20}
\end{align}
The last two equations show that our SU(2)-projected, U(1) potential possesses a
total-derivative \CSt.  Once we have in hand a \pr\ for~$A$ such that $A\rd A$ is a
total derivative, it is easy to find the \Cpr\ for~$A$. In the above,
\numeq{eq21}{
A = \rd{(-2\Phi)} + 2\bigl(1-(\sin^2\fract\omega2)\sin^2\Theta  \bigr) 
\rd{\bigl(\Phi + \tan^{-1}[ (\tan\fract\omega2) \cos\Theta]  \bigr)}\ .
}
The projected formula \refeq{eq18}, \refeq{eq21} for~$A$,
contains three arbitrary functions $\omega$, $\Theta$, and $\Phi$; this offers
sufficient generality to parameterize an arbitrary 3-vector~$\vec A$. Moreover, in
spite of the total derivative expression for $A\rd A$, its spatial integral need not
vanish. In our example, the functions $\omega$, $\Theta$, and $\Phi$ in
general depend on~$\vec r$; however, if we take~$\omega$ to be a function only of
$r =|\vec r|$, and identify 
 $\Theta$  and $\Phi$ with the polar and azimuthal angles  
$\theta$  and~$\phi$ of~$\vec r$, then 
\begin{align}
\int A\rd A &= 4\pi \int_0^\infty \rd r \frac{\rd{}}{\rd r}
(\omega-\sin\omega)\nonumber\\
&=  4\pi (\omega-\sin\omega) \Bigr|_{r=0}^{r=\infty}\ .
\label{eq22}
\end{align}
Thus if $\omega(0)=0$ and $\omega(\infty)= \pi N$, $N$ an integer, the integral is
nonvanishing, giving $4\pi^2N$; the contribution comes entirely from the bounding
surface at infinity~\cite{ref13}. 

With this preparation, I can now describe the \nA\ construction~\cite{ref14}. We are
addressing the following mathematical problem: We wish to parameterize a \nA\
vector potential $A^a$ belonging to a group~$H$, so that the \nA\ \CSt~\refeq{eq8} is
a total derivative. Since we are in three dimensions, the vector potential has
$3\times (\dim H)$ components, so our \pr\ should have that many
arbitrary functions.

The solution to our mathematical problem is to choose a large group~$G$
(compact,  semi\-simple) that contains~$H$ as a subgroup. The generators
of~$H$ are called $I^m$ [$m=1,\ldots$, $(\dim H)$] while those of $G$ not
in~$H$ are called $S^A$  [$A=1,\ldots, (\dim G) - (\dim H)$].
We further demand that
$G/H$ is a symmetric space; that is, the structure of the Lie algebra is 
\begin{subequations}\label{eq23}
\begin{align}
[I^m, I^n] &= f^{mno} I^o \label{eq23a}\\
[I^m, S^A] &= h^{mAB} S^B \label{eq23b}\\
[S^A, S^B] &\propto h^{mAB} I^m\ . \label{eq23c}
\end{align}
\end{subequations}
Here $f^{mno}$ are the structure constants of~$H$. Eq.~\refeq{eq23b} shows that the
$S^A$ provide a representation for~$I^m$ and, according to~\refeq{eq23c}, their
commutator closes on~$I^m$. The normalization of the $H$-generators is fixed by $\tr
I^m I^n = -N\delta^{mn}$. With~$g$, a generic group element of~$G$, giving rise to a
pure gauge potential $\mathcal A = \gdg$ in~$G$, we define the $H$-vector
potential~$A$ by projecting with generators belonging to~$H$:
\numeq{eq24}{
A = \frac1N \tr I^m \gdg\ .
}

We see that the Abelian [U(1)] construction presented in \refeq{eq12}--\refeq{eq16}
follows the above pattern: $\text{SU(2)}= G\supset H= \text{U(1)}$; $I^m=
\sigma^3/2i$, $S^A=\sigma^2/2i,\sigma^3/2i $. Moreover, a chain of equations
analogous to \refeq{eq14}--\refeq{eq17} shows that the $H$ \CSt\ is proportional to
$\tr(\gdg)^3$, which is  a total derivative~\cite{ref9,ref13}:
\numeq{eq30}{
\text{CS}(A\in  H) = \frac1{48\pi^2N} \tr(\gdg)^3\ .
} 

Two comments elaborate on our result. It may be useful to choose for~$H$ a direct
product $H_1\otimes H_2\subset G$, where it has already been established that the
\CSt\ of $H_2$ is a total derivative, and one wants to prove the same for the $H_1$
\CSt. The result~\refeq{eq30} implies that
\numeq{eq31}{
\text{CS}(A\in H_1) + \text{CS}(A\in H_2) = \frac1{48\pi^2N} \tr(\gdg)^3\ .
}
Since the right side is known to be a total derivative, and the second term on the left
side is also a total derivative by hypothesis, Eq.~\refeq{eq31} implies the desired
result that $\text{CS}(A\in H_1)$ is  a total derivative. Furthermore, since the total
derivative property of $\tr(\gdg)^3$ is not explicitly evident, our ``total derivative''
construction for a \nA\ \CSt\ may in fact result in an expression of the form $a\rd
a$, where~$a$ is an Abelian potential. At this stage one can appeal to known
properties of an Abelian \CSt\ to cast $a\rd a$ into total derivative form, for
example, by employing a \Cpr\ for~$a$. In other words, our construction may be
more accurately described as an ``Abelianization'' of a \nA\ \CSt.

To illustrate explicitly the workings of this construction, I present now the \pr\ for
an SU(2) potential $A_i= A_i^a \sigma^a/2i$, which contains $3\times3=9$ functions
in three dimensions. For~$G$ we take O(5), while $H$ is chosen as
$\text{O(3)}\otimes \text{O(2)} \approx \text{SU(2)} \otimes \text{U(1)}$, and we
already know that an Abelian [U(1)] \CSt\ is a total derivative. We employ a
4-dimensional representation for O(5) and take the $\text{O(2)}\approx \text{U(1)}$
generator to be $I^0$:
\begin{subequations}\label{eq32}
\numeq{eq32a}{
I^0 = \frac1{2i} \begin{pmatrix}
-I  & 0\\
0 & I 
\end{pmatrix}
}
while the $\text{O(3)} \approx \text{SU(2)}$ generators are $I^m$, $m=1,2,3$:
\numeq{eq32b}{
I^m = \frac1{2i} \begin{pmatrix}
\sigma^m & 0\\
0 & \sigma^m
\end{pmatrix}\ .
}
Finally, the complementary generators of O(5), which do  not belong to~$H$, are
$S^A$ and $\tilde S^A$, $A=1,2,3$:
\numeq{eq32c}{
S^A = \frac1{i\sqrt2}\begin{pmatrix}
0 & 0\\
\sigma^A & 0
\end{pmatrix}\ ,\quad 
\tilde S^A = \frac1{i\sqrt2}\begin{pmatrix}
0 & \sigma^A\\
0 & 0
\end{pmatrix}\ .
}
\end{subequations}
There are a total of ten generators, which is the dimension of O(5), and one verifies
that their Lie algebra is as in~\refeq{eq23}.

Next we construct a generic O(5)  group element~$g$, which is a $4\times4$ matrix.
The construction begins by choosing a special O(5) matrix~$M$, depending on six
functions, a generic O(3) matrix~$h$ with three functions, and a generic O(2)
matrix~$k$ involving a single function, for a total of ten functions,
\numeq{eq33}{
g = Mhk
}
where $M$ is given by 
\numeq{eq34}{
M = \frac1{\sqrt{1+\vec\omega\cdot\vec\omega^* -
\frac14(\vec\omega\times\vec\omega^*)}} 
\begin{pmatrix}
1-\frac i2 (\vec\omega\times\vec\omega^*)\cdot\vec\sigma &
-\vec\omega\cdot\vec\sigma \\
 \vec\omega^*\cdot\vec\sigma & 1 + \frac i2
(\vec\omega\times\vec\omega^*)\cdot\vec\sigma 
\end{pmatrix}\ .
}
Here $\vec\omega$ is a complex 3-vector, involving six arbitrary functions. The
SU(2) connection is now taken as in~\refeq{eq24}
\begin{subequations}\label{eq35}
\numeq{eq35a}{
A^m = -\tr (I^m \gdg)
}
and with~\refeq{eq33} this becomes
\begin{align}
A &= h^{-1} \tilde A h + h^{-1}\rd h\label{eq35b}\\
\tilde A &= -\tr (I^m M^{-1} \rd M)\ .  \label{eq35c}
\end{align}
\end{subequations}
We see that $k$ disappears from the formula for~$A$, which is an SU(2)
gauge-transform (with~$h$) of the connection~$\tilde A$ that is constructed just
from~$M$. It is evident that~$A$ depends on the required nine parameters: three
in~$h$ and six in~$M$. 

[Interestingly, the \pr\ \refeq{eq35} of the SU(2) connection possess a structure
analogous to the \Cpr\ of an Abelian vector. Both present their connection as a gauge
transformation of another, ``core'' connection: $\theta$ in the Abelian formula
$\grad\theta + \alpha\grad\beta$, and $h$ in~\refeq{eq35b}.]

The \CSt\  \refeq{eq35b} of $A$ in \refeq{eq35a} relates to that of \refeq{eq35c} by
a gauge transformation:
\numeq{eq36}{
\text{CS}(A) = \text{CS}(\tilde A) + \rd{\tr\Bigl(-\frac1{8\pi^2} h^{-1}\rd h \tilde A
\Bigr)} + \frac1{48\pi^2} \tr (h^{-1}\rd h)^3 \ .
 }
The last two terms on the right describe the response of a \CSt\ to a gauge
transformation; the next-to-last is manifestly a total derivative, as is the last  --
in a ``hidden'' fashion. Finally,
\numeq{eq37}{
\text{CS}(\tilde A) = \frac1{16\pi^2} a\rd a 
}
where
\numeq{eq38}{
a = \frac{\vec\omega\cdot \rd{\vec\omega^*}- \,\vec\omega^*\cdot\rd\omega}{1
+\vec\omega\cdot\vec\omega^* - \frac14(\vec\omega\times\vec\omega^*)^2}
 }
We remark that $a$ can now be \prd\ in the Clebsch manner, so that $a\rd a$
appears as a total derivative, completing our construction.

\end{document}